\documentclass[doublecol]{epl2} 
\usepackage{amssymb}

\title{Cosmological Tests with the Joint Lightcurve Analysis}
\shorttitle{Type Ia Supernovae} 

\author{F. Melia\inst{1}\thanks{John Woodruff Simpson Fellow.},
J.-J. Wei\inst{2}, R.~S. Maier\inst{3}, and X.-F. Wu\inst{4}}
\shortauthor{F. Melia, J.-J. Wei, R. S. Maier and X.-F. Wu}

\institute{
  \inst{1}Depts.\ of Physics and Astronomy, and The Applied Math Program, 
The University of Arizona, AZ 85721 USA\\ \null\hskip 0.12in\email{fmelia@email.arizona.edu} \\
  \inst{2}Purple Mountain Observatory, Chinese Academy of Sciences, Nanjing 210008, China \\ \null\hskip 0.12in\email{jjwei@pmo.ac.cn}\\
  \inst{3}Depts.\ of Mathematics and Physics, and The Statistics Program, 
The University of Arizona, AZ 85721 USA\\ \null\hskip 0.12in\email{rsm@math.arizona.edu} \\
  \inst{4}Purple Mountain Observatory, Nanjing 210008, China; \email{xfwu@pmo.ac.cn}
}


\abstract{We examine whether a comparison between 
$w$CDM and $R_{\rm h}=ct$ using merged Type~Ia SN catalogs produces 
results consistent with those based on a single homogeneous sample. 
Using the Betoule et~al.\ \cite{Betoule:2014} joint
lightcurve analysis (JLA) of a combined sample of 613 events from SNLS
and SDSS-II, we estimate the parameters of the two models and compare
them. We find that the improved statistics can alter the model
selection in some cases, but not others. In addition, based on the model
fits, we find that there appears to be a lingering systematic offset of
$\sim 0.04$--$0.08$ mag between the SNLS and SDSS-II sources,
in spite of the cross-calibration in the JLA\null. Treating $w$CDM, $\Lambda$CDM
and $R_{\rm h}=ct$ as separate models, we find in an unbiased pairwise 
statistical comparison that the Bayes Information Criterion (BIC) favors the 
$R_{\rm h}=ct$ Universe with a likelihood of $82.8\%$ versus $17.2\%$
for $w$CDM, but the ratio of likelihoods is reversed ($16.2\%$ versus 
$83.8\%$) when $w_{\rm de}=-1$ (i.e., $\Lambda$CDM) and strongly 
reversed ($1.0\%$ versus $99.0\%$) if in addition $k=0$ (i.e., flat
$\Lambda$CDM). We point out, however, that the value of $k$ is 
a measure of the net energy (kinetic plus gravitational) in the Universe
and is not constrained theoretically, though some models of inflation would
drive $k\rightarrow 0$ due to an expansion-enforced dilution. Since we here
consider only the basic $\Lambda$CDM model, the value of $k$ needs to be 
measured and, therefore, the pre-assumption of flatness introduces a 
significant bias into the BIC.}  

\begin{document}

\maketitle

\section{Introduction}
\label{sec:1}
A study of the Type~Ia SN Hubble diagram has allowed us to measure the
expansion history of the Universe
\cite{Perlmutter:1998,Riess:1998,Schmidt:1998}.  But as successful as
this program has been, it relies on the use of integrated quantities
that are not independent of the assumed dynamics.  The use of SN
measurements in conducting unbiased, comparative studies of
alternative expansion histories is therefore intricate, because
at~least three `nuisance' parameters characterizing the standard
candle must be optimized simultaneously with any cosmological model's
free parameters \cite{Suzuki:2012,Melia:2012a}.

It has therefore been useful to seek additional methods of probing the
cosmic spacetime, including the analysis of cosmic chronometers
\cite{MeliaMaier:2013}, gamma-ray bursts \cite{Wei:2013},
high-$z$ quasars \cite{Melia:2013a}, the cosmic microwave
background \cite{Melia:2014}, and Baryon Acoustic
Oscillations (BAO)
\cite{Delubac:2014}.  However, the
results of these studies differ from the Type~Ia SN perception that
either $w$CDM (with dynamical dark energy) or $\Lambda$CDM (with a 
cosmological constant)  is the optimal cosmological model.  Instead, they tend to favor
another Friedmann--Robertson--Walker (FRW) cosmology known as the
$R_{\rm h}=ct$ Universe
\cite{Melia:2007,MeliaShevchuk:2012,Melia:2016,Melia:2017}.  For example,
the most recent application of the Alcock--Paczy\'nski test to
model-independent BAO data
\cite{Delubac:2014,MeliaLopez:2017} has favored $R_{\rm
  h}=ct$ over $w$CDM at better than a $99.34\%$ confidence level
\cite{MeliaLopez:2017}.

It is therefore desirable to compare $w$CDM, $\Lambda$CDM and 
$R_{\rm h}=ct$ directly, using the SN measurements themselves. In our 
previous paper \cite{Wei:2015}, we carried out such a comparative analysis
based on a single, homogeneous Type~Ia SN sample---the Supernova
Legacy Survey (SNLS; ref.~\cite{Guy:2010}).  It is well known now that
merging different subsamples, each with its own set of possibly
unknown systematics, can introduce inconsistencies that reduce the
power of Type~Ia SNe for model comparisons.  Various techniques have
been employed to address this problem, including the introduction of
an intrinsic dispersion for each subsample, constrained by the
requirement that the $\chi^2_{\rm dof}$ of the fit be equal to one in
each case \cite{Amanullah:2010,Suzuki:2012} (See also the
general discussion in ref.~\cite{Betoule:2014}.) It is questionable 
whether this approach provides a statistically fair selection between 
different models. In our previous analysis using a single large compilation, 
rather than a merger of unrelated subsamples, the results were quite 
clear: In a pairwise comparison, the Bayes Information Criterion (BIC)
favors $R_{\rm h}=ct$ over $\Lambda$CDM with a likelihood of $\approx 90\%$
versus only $\approx 10\%$ for the standard model. The ratio of
likelihoods is even greater when comparing $R_{\rm h}=ct$ with $w$CDM.
 
Recent progress was made using merged samples \cite{Betoule:2014} 
by introducing a new cross-calibration of the Sloan Digital Sky Survey (SDSS-II;
$0.05<z<0.4$) \cite{Sako:2015} and SNLS ($0.2<z<1$) \cite{Guy:2010}
samples. In this \emph{Letter}, we repeat our one-on-one comparison of $w$CDM
and $\Lambda$CDM with $R_{\rm h}=ct$, though this time using the joint analysis 
of these SDSS-II and SNLS sub-samples \cite{Betoule:2014} with 613 SNe~Ia.  
We examine whether the outcome of such a study using merged
samples is consistent with that based solely on a single (though
relatively large) compilation.

\section{The Combined Supernova Sample}
\label{sec:2}
The Union2.1 catalog \cite{Suzuki:2012}, which currently includes $580$
SN detections \cite{Kuznetsova:2008,Riess:2011}, is a merger of subsamples,
each with its own set of systematic and intrinsic uncertainties, commonly 
subsumed into an unknown \emph{intrinsic} dispersion~$\sigma_{\rm int}$. 
Note, however, that
some systematic uncertainties may relate to the Hubble diagram as a
whole, in which case they do not contribute to $\sigma_{\rm int}$.
For this \emph{Letter}, we do not have sufficient information to separate
the two. For this reason, and the fact that the cross-calibration
in the JLA allowed ref.~\cite{Betoule:2014} to find a a single sample-wide 
dispersion in place of individual $\sigma_{\rm int}$'s, we shall also adopt
their approach here. In other circumstances, however, where a
cross-calibration is not available \cite{DAgostini:2005,Kim:2011,Wei:2015},
it is questionable whether model parameters can
be estimated by minimizing an overall~$\chi^2$ (while constraining the
$\chi^2_{\rm dof}$ of each subsample to equal unity), since the
unknown $\sigma_{\rm int}$'s should be estimated simultaneously with
all other parameters \cite{Kim:2011}. As an alternative, the method of
maximum likelihood estimation (MLE) has been shown to yield superior
results \cite{DAgostini:2005,Kim:2011,Wei:2015}, though when multiple
$\sigma_{\rm int}$'s are used, the analysis becomes computationally
challenging. We shall see that, even though we shalll adopt the 
single sample-wide dispersion of ref.~\cite{Betoule:2014}, MLE is still 
required because the probabilities depend on nuisance parameters
that need to be optimized with the fits (see discussion following
Eq.~(\ref{eq:6}) below).

Some catalogs available for this work are large and well 
suited to the analysis we carry~out in this \emph{Letter}.
For example, about half of the Type~Ia SNe in Union2.1 came from the
SNLS \cite{Guy:2010}.  Since the same instruments and reduction
techniques were employed for all 252 of these ($0.15<z<1.1$)
events, a single $\sigma_{\rm int}$ is sufficient to characterize the
unknown intrinsic scatter.

In this \emph{Letter}, we optimize the cosmological fits
using the joint lightcurve analysis (JLA) \cite{Betoule:2014}, 
a true recalibration of all the data in SNLS and SDSS-II, based on the 
use of tertiary standard stars observed by both experiments as
reference, a common point-spread function, and the same procedural
steps, yielding a photometric accuracy approaching $\approx 5$~mmag
(but see the discussion below of a possible relic offset between them). 
The merged sample has a consistent calibration and systematics. 

Two features of the JLA catalog need to be addressed, however.
First, the complete catalog contains not only the SNLS and
SDSS-II events used in the cross-calibration, but also an
additional low-$z$ ($z< 0.1$) sample, and the
HST SNe at high redshift ($z\approx 1$).  Neither of these two
groupings was involved in the cross-calibration, so the problem of
disparate systematics and intrinsic differences remains for them. The
low-$z$ events were calibrated against secondary
photometric standards \cite{Landolt:2007}, but one must allow
for measurement uncertainties in the reference F-subdwarf
BD +17~4708 system used there.  The calibration of the HST SNe was
based on the Riess et~al.\ \cite{Riess:2007} interpretation. Consistent
with our goal of avoiding unknown systematics and calibration
uncertainties as much as possible, we shall not include
these two (small) subsamples in our analysis because their calibration
was handled differently from that of SNLS and
SDSS-II\null.  

Second, ref.~\cite{Betoule:2014}
managed to avoid some of the problems associated with the introduction
of intrinsic dispersions, $\sigma_{\rm int}$, but could not completely
eliminate them. However, instead of treating them as `nuisance'
parameters to be estimated during the analysis itself (as was done
previously), they attempted to find a model-independent dispersion as
a function of redshift, by partitioning the SNe into different
redshift intervals and using the distribution of observed magnitudes
to guess an overall dispersion in each bin. In their assessment, the
various dispersions `measured' in this way are consistent with a
single, constant value $\sigma_{\rm coh}=\allowbreak 0.106\pm0.006$
(labeled in this fashion to distinguish it from the older~$\sigma_{\rm
  int}$'s). This is the sample-wide `intrinsic' dispersion we shall
also use for the JLA in this \emph{Letter}, consistent with the approach of
ref.~\cite{Betoule:2014}.

The distance modulus of a given supernova is inferred from a fit of
its spectral evolution using one of several light-curve models. 
For this step, two methods are commonly used,
SiFTO and SALT2~\cite{Conley:2008}. Ref.~\cite{Betoule:2014}
computed the (observed) distance modulus of each Type~Ia SN using 
SALT2, and since we are closely following their approach, we shall
also use this lightcurve fitter in our analysis. An argument in favour
of this choice for the JLA is that SALT2 is data-driven, and does not
introduce any significant bias between low and high-redshift distances
\cite{Mosher:2014}, which is essential when dealing with a large
sample that spans the redshift range we have here. SALT2 includes 
the following components: (i)~the apparent
magnitude, $m_B$, of the SN at maximum light; (ii)~the `shape,'
$X_{1}$, of its lightcurve; and (iii)~its deviation,~$C$, from the
mean Type~Ia SN $B-V$ color. The formula for the distance modulus is
$\mu_{B} \equiv\allowbreak m_{B}+\alpha\cdot\nobreak X_{1}-
\beta\cdot\nobreak C-\nobreak M_{B}$, where $M_{B}$~is the absolute
{\it B}-band magnitude of a Type~Ia SN with $X_{1}=0$
and~$C=0$. An allowance may also be made for the assumed host galaxy
mass
\cite{Lampeitl:2010,Suzuki:2012},
introduced as an adjustment $\Delta M_{\rm host}$ to~$M_{B}$ for SNe
in host galaxies with a mass $>10^{10}M_\odot$ \cite{Betoule:2014}.

`Nuisance' parameters used in computing each distance modulus,
such as $\alpha$, $\beta$, $M_B$ (and $\Delta M_{\rm host}$ if
present), must be fitted simultaneously with parameters
characterizing the cosmology itself.  Many
previously reported model comparisons have not uniformly recalibrated
the data for each cosmology being tested. For example,
in ref.~\cite{Melia:2012a} we adopted the lightcurve parameters
optimized for $\Lambda$CDM and used them for all the models being
tested. But this simplifying assumption can vitiate the
outcome of model selection. One goal of this 
\emph{Letter} is to relax this restriction and re-optimize the nuisance
parameters separately for each model. When this procedure is
followed, past experience has shown that the cosmology preferred 
by the Type~Ia SN data is not always $w$CDM or $\Lambda$CDM
\cite{Wei:2015}.

\section{Model Comparisons}
\label{sec:3}
For each SN, the theoretical distance modulus $\mu_{\rm th}$ is
calculated from the measured redshift~$z$ by the definition $\mu_{\rm
  th}(z)\equiv5\log[D_{\rm L}(z)/ 10\; {\rm pc}]$, where $D_{\rm
  L}(z)$ is the model-dependent luminosity distance. $w$CDM 
and $\Lambda$CDM assume specific constituents in the density, expressed as
$\rho=\allowbreak \rho_{\rm r}+ \nobreak \rho_{\rm m}+\nobreak
\rho_{\rm de}$, where $\rho_{\rm r}$, $\rho_{\rm m}$ and~$\rho_{\rm
  de}$ are, respectively, the energy densities of radiation, (luminous
and dark) matter, and dark energy.  These densities are often
represented in~terms of today's critical density, $\rho_{\rm c}\equiv
3c^2 H_0^2/8\pi G$, as $\Omega_{\rm m}\equiv\rho_{\rm m}/\rho_{\rm
  c}$, $\Omega_{\rm r}\equiv\rho_{\rm r}/\rho_{\rm c}$, and
$\Omega_{\rm de}\equiv \rho_{\rm de}/\rho_{\rm c}$.  $H_0$~is the
Hubble constant, and the other symbols have their usual meanings.  In
$R_{\rm h}=ct$, on the other hand, whatever constituents are present
in~$\rho$ beyond matter and radiation, the principal constraint is the
zero active mass condition
\cite{Melia:2007,MeliaShevchuk:2012,Melia:2016,Melia:2017}, which
corresponds to a \emph{total} equation-of-state $p=-\rho/3$.

$w$CDM has a dark energy with an equation-of-state 
$p_{\rm de}=w_{\rm de}\,\rho_{\rm de}$ and $w_{\rm de}\not=-1$. 
Its luminosity distance is given by the expression
\begin{eqnarray}
D_{\rm L}^{w{\rm CDM}}(z)&=&\frac{c}{H_{0}}\,\frac{(1+z_{\rm hel})}{\sqrt{\mid\Omega_{k}\mid}}
\,{\rm sinn}\nonumber\\
&\null&\hskip -1.1in\left\{\int_{0}^{z}\frac{{\rm d}z\;\mid\Omega_{k}\mid^{1/2}}
{\sqrt{\Omega_{\rm m}(1+z)^{3}+\Omega_{k}(1+z)^{2}+\Omega_{\rm de}(1+z)^{3(1+w_{\rm de})}}}\right\}\;\;\;\;\;
\label{eq:3}
\end{eqnarray}
where $z$ and $z_{\rm hel}$ are the CMB rest frame and heliocentric
redshifts of the SN, and $\Omega_{k}=\allowbreak 1-\nobreak
\Omega_{\rm m}- \nobreak \Omega_{\rm de}$ represents the spatial
curvature of the Universe---appearing as a term proportional to~$k$ in
the Friedmann equation. In addition, $\rm sinn$ is $\sinh$ when
$\Omega_{k}>0$ and $\sin$ when $\Omega_{k}<0$.  For a flat Universe
with $\Omega_{k}=0$, the right-hand side of this expression simplifies
to the form $(1+\nobreak z_{\rm hel}) c/H_{0}$ times the integral
(though without the $\Omega_{k}$ term).  

Depending on the application,
the standard model may contain as many as ten parameters, though only
three of these are critical for supernova work. One may adjust
$\Omega_{\rm m}$, $k$ (or equivalently~$\Omega_{\rm de}$), and~$w_{\rm
  de}$. It~is well known that $H_0$~is degenerate with~$M_B$ when
constructing a SN Hubble diagram, so it is not free if $M_B$~is one of
the optimized variables \cite{Suzuki:2012}. 

If dark energy is assumed to be a cosmological constant, 
with $w_{\rm de}=-1$, the standard model becomes $\Lambda$CDM\null.
The principal parameters in this 
model are $\Omega_{\rm m}$ and the spatial curvature constant~$k$.
One often sees flatness (i.e., $k=0$) assumed on the basis of
other kinds of observation, but there are good reasons to avoid
this if possible. Unlike the distinction between $w_{\rm de}\not=-1$,
which represents dynamical dark energy, and $w_{\rm de}=-1$, which
represents a cosmological constant, there is no theoretical basis to
distinguish $k=0$ and $k\not=0$, though some models of
inflation would have driven $k\rightarrow 0$ via an expansion-enforced 
dilution of the Universe's net energy (kinetic plus gravitational).
But since we are here considering only a basic $\Lambda$CDM model,
the value of~$k$ is an initial condition, not a fundamental constraint, 
and must be measured from the observations, as is routinely
done, e.g., with anisotropies in the cosmic microwave background. When 
one uses parameters optimized in previous studies, however, one is obliged 
to use other parameters optimized in correspondence with these values, 
which would actually yield less favorable fits to the SN data. 
In other words, when comparing models,
it is not fair statistically to adopt a previously optimized value of~$k$, 
while ignoring other parameters that are then re-optimized with SN data. 
The two models we have at our disposal for unbiased SN work are therefore
$w$CDM, with three free parameters: $\Omega_{\rm m}$, $\Omega_{\rm de}$, 
and~$w_{\rm de}$, and $\Lambda$CDM with $\Omega_{\rm m}$ and
$\Omega_{\rm de}\equiv \rho_\Lambda/\rho_{\rm c}$. But to demonstrate
how critical this issue of pre-optimized parameters can be, we shall also
show the result of model comparisons using flat $\Lambda$CDM\null.
As we shall see, the JLA sample is so large now that even one change 
in the handling of the parameters in the standard model can greatly 
alter the outcome of the analysis.

The luminosity distance in $R_{\rm h}=ct$ is given by the simpler
expression
\begin{equation}
D_{\rm L}^{R_{\rm h}=ct}(z)=\frac{c}{H_{0}}(1+z_{\rm hel})\ln(1+z)\;.
\label{eq:4}
\end{equation}
Since $H_0$ is degenerate with $M_B$, the $R_{\rm h}=ct$ Universe has
no parameters to adjust when we construct its SN Hubble diagram.
Further discussion on observational differences between $w$CDM and
$R_{\rm h}=ct$ appears in
refs.\ \cite{Melia:2007,Melia:2012a,Melia:2013b,MeliaShevchuk:2012,MeliaMaier:2013,Wei:2013}.
For a pedagogical treatment, see also ref.~\cite{Melia:2012b}.

The cosmological parameters, along with the model-specific nuisance
parameters, are estimated using an approach first described in
refs.~\cite{DAgostini:2005,Kim:2011}, and more fully developed by us in
ref.~\cite{Wei:2015}. It is based on the joint likelihood function to
be maximized for all these parameters, or as a multiplicative factor 
that modifies an assumed flat Bayesian prior.  This function is
\begin{equation}
L = \frac{\exp\left[-\frac{1}{2}( \bf{\hat{\mu}_{B}}-\bf{\hat{\mu}_{\rm th}} )^{T}\;
    \textbf{C}^{-1}\; ( \bf{\hat{\mu}_{B}}-\bf{\hat{\mu}_{\rm th}} ) \right]}{\sqrt{(2\pi)^{n}\det \textbf{C}}}\;,
\label{eq:5}
\end{equation}
where $\bf{\hat{\mu}_{B}}$ ($\bf{\hat{\mu}_{\rm th}}$) is the observed
(theoretical) distance modulus vector with $n$~components, $n$~being
the number of SNe, and $\textbf{C}$~is the full $n\times
n$ covariance matrix (including both statistical and systematic
errors), defined by
\begin{equation}
\textbf{C}=\textbf{D}_{\rm stat}+\textbf{C}_{\rm stat}+\textbf{C}_{\rm sys}\;.
\end{equation}
In this, $\textbf{D}_{\rm stat}$ is the diagonal part of the statistical
uncertainty, given by
\begin{eqnarray}
(\textbf{D}_{\rm stat})_{ii}&\hskip-0.1in=\hskip-0.1in&
\sigma^{2}_{m_{B},i}+\alpha^{2}\sigma^{2}_{X_{1},i}+\beta^{2}\sigma^{2}_{\mathcal{C},i}\nonumber\\*
&\hskip-0.1in\null\hskip-0.1in&+C_{m_{B}\,X_{1}\,\mathcal{C},i}
+\sigma^{2}_{{\rm pec},i}+\sigma^{2}_{{\rm lens},i}+\sigma^{2}_{\rm coh}\;,
\end{eqnarray}
where $\sigma_{{m_{B}},i}$, $\sigma_{{X_{1}},i}$, and
$\sigma_{{\mathcal{C}},i}$ are the standard errors of the peak
magnitude and light-curve parameters of the $i$'th~SN\null. The term
$C_{m_{B}\,X_{1}\,\mathcal{C},i}$ derives from the covariances among
$m_{B},X_{1},\mathcal{C}$, and itself depends quadratically on the
nuisance parameters $\alpha,\beta$.  The dispersion $\sigma_{{\rm
    pec},i}=\allowbreak 5\sigma_z/(z_i\log{10})$ accounts for the
uncertainty in cosmological redshift due to peculiar velocities, and
$\sigma_{{\rm lens},i}$ accounts for the variation of magnitudes
caused by gravitational lensing.  We follow ref.~\cite{Betoule:2014} 
in using $c\sigma_z=150$ km~s$^{-1}$, as
well as $\sigma_{{\rm lens},i}=0.055\times z_i$, as suggested in
ref.~\cite{Jonsson:2010}.  The statistical and systematic covariance
matrices, $\textbf{C}_{\rm stat}$ and $\textbf{C}_{\rm sys}$, are
generally not diagonal \cite{Conley:2011}, and for the JLA are given by
\begin{eqnarray}
\textbf{C}_{\rm stat}+\textbf{C}_{\rm sys}&\hskip-0.1in=\hskip-0.1in&
\textbf{V}_{0}+\alpha^{2}\textbf{V}_{a}+\nonumber\\
&\hskip-0.1in\null\hskip-0.1in&\hskip-0.4in\beta^{2}\textbf{V}_{b}
+2\alpha \textbf{V}_{0a}-2\beta \textbf{V}_{0b}-2\alpha \beta \textbf{V}_{ab}\;,
\label{eq:6}
\end{eqnarray}
where $\textbf{V}_{0}$, $\textbf{V}_{a}$, $\textbf{V}_{b}$,
$\textbf{V}_{0a}$, $\textbf{V}_{0b}$, and $\textbf{V}_{ab}$ are
available from ref.~\cite{Betoule:2014} at
$\underline{\rm {http://supernovae.in2p3.fr}}$.  From the resulting
model-specific likelihood function~$L$, which can also be viewed as
a Bayesian posterior, we determine the best-fit values for the
parameters by maximizing over the joint parameter space. 
In this just-described estimation, we take into account the 
extensive analysis carried out in refs.~\cite{Conley:2011,Betoule:2014}, 
concerning systematic errors.

Each distance modulus~$\mu_{B,i}$ depends on $\alpha,\beta,M_B$, and
$\Delta M_{\rm host}$.  As stated, we maximize the likelihood~$L$ over
all nuisance parameters.  Kim \cite{Kim:2011} notes that such a
`full~MLE' is better founded statistically, since it treats on the
same level all cosmological and all nuisance parameters, the
uncertainties in which can affect each other, including those of the
intrinsic dispersion(s), if these are themselves not known a
priori. (This will not be the case here, since $\sigma_{\rm coh}$
is fixed at~$0.106$. However, we do not expect the lack of
individual optimization of $\sigma_{\rm coh}$ for each model to
significantly affect our results~\cite{Rubin:2015}.)
We emphasize, however, that even though $\sigma_{\rm coh}$
is fixed in this approximation, the Gaussian normalization in our likelihood 
analysis is still not a constant. It depends on the value of $\alpha$, $\beta$, $M_B$, and
$\Delta M_{\rm host}$ (see Eqn.~5). Thus, maximizing the
likelihood function $L$ is not exactly equivalent to minimizing the 
$\chi^{2}$ statistic, i.e., 
$\chi^{2}=( \bf{\hat{\mu}_{B}}-\bf{\hat{\mu}_{\rm th}} )^{T}\;\textbf{C}^{-1}\; ( \bf{\hat{\mu}_{B}}-\bf{\hat{\mu}_{\rm th}} )$.

In addition to computing the best-fit parameter values from the
likelihood function~$L$ by MLE, we treat~$L$ in a Bayesian
  fashion as an unnormalized probability density function (PDF) on
the joint parameter space, and employ Markov-chain Monte Carlo (MCMC)
techniques to generate a large random sample of points from this
space, distributed according to the~PDF\null.  The standard error
  for each estimated parameter is then obtained as an empirical
  standard deviation of this sample.

Because $w$CDM, $\Lambda$CDM and $R_{\rm h}=ct$ have different
numbers of free parameters, comparing their likelihoods of being the 
`correct' model requires the use of a model selection criterion.  Since the samples 
we are dealing with here are very large, the most appropriate tool to use
\cite{MeliaMaier:2013} is the Bayes Information Criterion, which
approximates the computation of the (logarithm of the) `Bayes factor'
for deciding between models \cite{Schwarz:1978,Kass:1995}.
The BIC is defined for each model being fit by
\begin{equation}
\label{eq:7}
\exp(-{\rm BIC}/2)\equiv n^{-p/2} L^*\;,
\end{equation}
where $L^*$ is the maximized likelihood, $n$ ($=613$ here) the
data set size, and $p$~the count of free parameters in the
model.  If ${\rm BIC}_\alpha$ comes from model~$\alpha$, the
unnormalized likelihood of model $\alpha$ being correct is the `Bayes
weight' $\exp(-{\rm BIC}_\alpha/2)$.  Thus model~$\alpha$
($\alpha=1,2$) has likelihood
\begin{equation}
P(\alpha)=
\frac{\exp(-{\rm BIC}_\alpha/2)}
{\exp(-{\rm BIC}_1/2)+\exp(-{\rm BIC}_2/2)}
\end{equation}
of being the correct choice. This has a Bayesian interpretation:
$\exp\left(-{\rm BIC}_\alpha/2\right)$ is a large-sample
($n\to\infty$) approximation to an integral over the parameter
space of model~$\alpha$, of its likelihood function~$L$.  In the
limit, the standard error of each parameter shrinks like~$n^{-1/2}$,
and the integral of~$L$ equals up~to a constant factor the
$n^{-p/2}L^*$ of Eq.~(\ref{eq:7}).

By convention, the magnitude of the difference $\Delta\equiv 
{\rm BIC}_2-{\rm BIC}_1$ provides a numerical assessment of the evidence
that model~$1$ is favoured over model~$2$. The rule of thumb is that
if $\Delta\lesssim 2$, the evidence is weak; if $\Delta\sim 3$ or~$4$, it
is mildly strong, and if $\Delta\gtrsim 5$, it is quite strong.

\section{Hubble Diagram}
\label{sec:4}
Several of the SNLS and SDSS-II supernovae fall outside
the range of validity established for the lightcurve fitter, SALT2,
and must therefore be removed \cite{Betoule:2014}. The pruned catalogs 
include 239 events from SNLS and 374 from SDSS-II, for a total of 613 
events. 

\begin{figure}
\onefigure[width=2.8in]{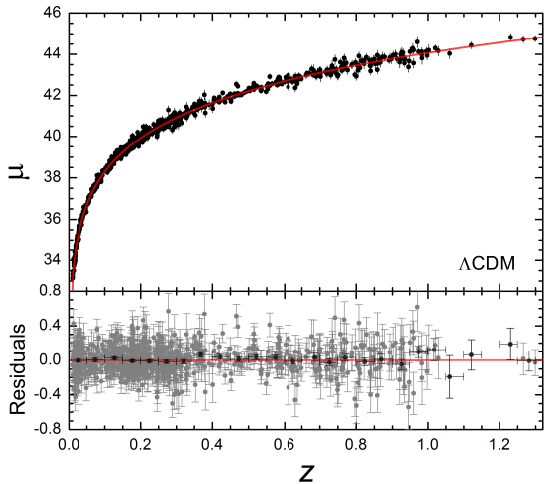}
\caption{\footnotesize (Top) Hubble diagram for the
  combined SNLS and SDSS-II sample of 613 Type~Ia SNe, together with
  the best-fit $w$CDM model (red curve) (line~2 in
  Tables 1 and~2)\null. (Bottom) Residuals for the best-fit model,
  shown in grey for individual sources, and in black for averages over
  redshift bins of~0.05.}
\label{fig:1}
\end{figure}

The best-fit parameters for $w$CDM, $\Lambda$CDM and $R_{\rm h}=ct$, 
obtained by MLE, are provided in Tables 1 and~2, along with a standard error
for each (estimated by MCMC\null).\footnote{A chain of $10^5$ points in
  the parameter space distributed according to the likelihood function
  was generated from the Metropolis--Hastings algorithm with a uniform
  prior.  In each case, the distributions of the estimated parameters, 
  with a confidence interval for each, followed from a statistical
  analysis.}  The first comparison we make is an extension to our previous 
work based solely on the SNLS~\cite{Wei:2015}.  The principal motivation 
in this \emph{Letter} has been to examine whether the outcome of that analysis 
is supported by a similar comparative study involving a much bigger, merged
sample. The corresponding results for the combined JLA sample are
summarized in lines 1, 2 and 3 of Tables 1 and~2 for $R_{\rm h}=ct$,
$w$CDM and $\Lambda$CDM\null. As indicated earlier, we also compare
these models with flat $\Lambda$CDM (line~4) to demonstrate the impact
of adopting a pre-optimized parameter value (for~$k$). The optimization for 
$R_{\rm h}=ct$, $w$CDM and $\Lambda$CDM is based purely on the SN 
observations.  

\begin{figure}
\onefigure[width=2.8in]{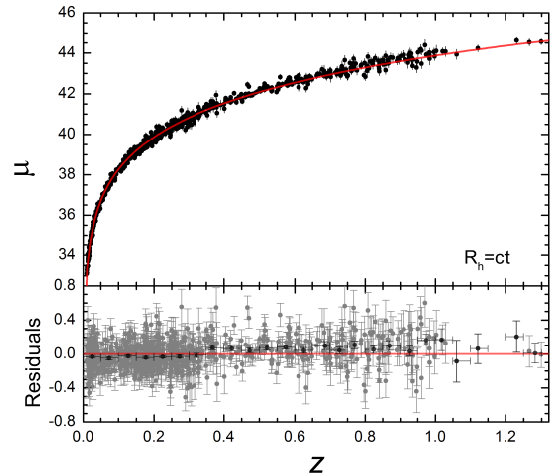}
\caption{Similar to Fig.~\ref{fig:1}, except that these
  data are calibrated using the $R_{\rm h}=ct$ Universe, and the red
  curve shows the best fit in this model (see line~1,
  Tables 1 and~2).}
\label{fig:2}
\end{figure}

On the basis of the JLA, the BIC favors $R_{\rm h}=ct$ over 
$w$CDM, with a likelihood of $82.8\%$ versus $17.2\%$. With $\Delta=3.14$,
the evidence in favour of $R_{\rm h}=ct$ is mildly strong. However,
both SNLS and the combined JLA sample are so big that the BIC (see
Eq.~(\ref{eq:7})) is sensitive to the number of free parameters. For
example, with one fewer parameter than $w$CDM, $\Lambda$CDM 
is somewhat favoured over $R_{\rm h}=ct$ with $\Delta=3.29$,
and strongly favoured over $w$CDM with $\Delta=6.43$. 
The sensitivity of this outcome to the various assumptions is further
demonstrated by the sample selection. Notice, for example, that
the SNLS on its own yields very different likelihoods. In this case,
even a comparison between $R_{\rm h}=ct$ and flat $\Lambda$CDM
shows that  the likelihoods are about even, i.e., $43.2\%$ versus 
$56.8\%$. The statistically fairer comparison between $R_{\rm h}=ct$
and $w$CDM and $\Lambda$CDM shows that the evidence in favour
of the former is mildly---or even very---strong in both cases. 

Yet in every case, the $\chi^2_{\rm dof}$ values for the best fit
models are hardly distinguishable. It is clear that 
model selection using Type~Ia SNe is therefore heavily influenced 
by the number of free parameters in the models. And given
this sensitivity, it is necessary to avoid biasing
the results by assigning pre-optimized values to the
variables that are not theoretically constrained (such as $k$
in this case).

\begin{figure}
\onefigure[width=2.9in]{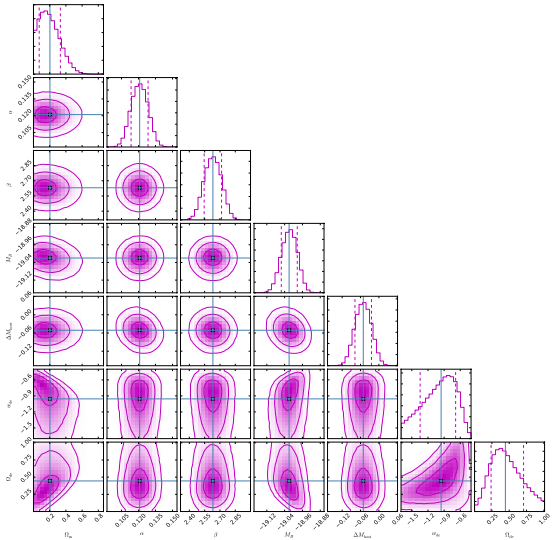}
\caption{Posterior probability distributions of the free parameters 
  in $w$CDM\null. Contours are 1,~2, and 3~$\sigma$. The vertical lines 
  are the best-fit results (solid), and the enclosed 
  $68\%$ credible region (dashed).  Made with
  \texttt{triangle.py} from ref.~\cite{Foreman:2013}.}
\label{fig:3}
\end{figure}

\begin{figure}
\onefigure[width=2.4in]{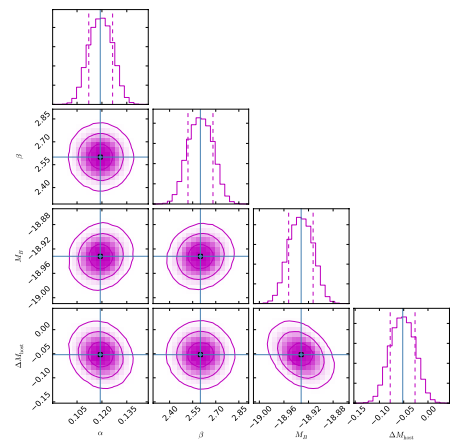}
\caption{Same as Fig.~\ref{fig:3}, but now for the
  $R_{\rm h}=ct$ Universe.}
\label{fig:4}
\end{figure}

The corresponding Hubble diagrams for the best-fit $w$CDM (line~2) and
$R_{\rm h}=ct$ (line~1) models are shown in Figs.~\ref{fig:1}
and~\ref{fig:2}, respectively, together with their residuals.  A close
inspection of the data in these plots shows that, though very similar,
they are not identical, highlighting the importance of estimating the
nuisance parameters individually for each different model. It is 
also quite evident from a comparison of the best-fit curves in these plots
that both models fit the data extremely well; the $\chi^2_{\rm dof}$
values attest to this, and demonstrate a comparably high quality fit
in each case.  We also show in Figs.~\ref{fig:3} 
and~\ref{fig:4} the corresponding one- and two-dimensional
projections of the posterior probability distributions for the free
parameters in $w$CDM and $R_{\rm h}=ct$, generated by~MCMC.

\section{Discussion and Conclusions}
\label{sec:5}
In this \emph{Letter}, we have used the MLE method (with
Bayesian extensions).  In our previous study of the SNLS sample
\cite{Wei:2015}, we also employed MLE, but in addition, contrasted the
outcome with that of the more conventional procedure of optimizing
$\sigma_{\rm int}$ \cite{Amanullah:2010,Suzuki:2012} by requiring that
$\chi^2_{\rm dof}$ equal unity. When using a single, homogeneous
sample, these two approaches give essentially the same result, because
(as we have seen) the $\chi^2_{\rm dof}$ of the optimized fit is
almost always close to~$1$. There is less justification \cite{Betoule:2014} 
for using the latter approach when several subsamples are merged into a 
single compilation.  Thus, even though the use of MLE with a full covariance
matrix is computationally difficult, it should be the method of choice
for any model selection involving non-nested models and a blend of
diverse subsamples.

The outcome of our analysis using the combined SNLS and
SDSS-II sample is in agreement with that of our earlier study based 
solely on the SNLS for some cases, but not others. The $R_{\rm h}=ct$
universe is consistently favoured over $w$CDM, but the likelihoods
are reversed when comparing $R_{\rm h}=ct$ with $\Lambda$CDM\null.
If we introduce a previously optimized value for~$k$, the outcome
is strongly reversed.

\begin{table*}
\centering
{
\tiny
\begin{tabular}{lccccc}
Model$\qquad$&$\quad${$\alpha$}$\qquad$&$\quad${$\beta$}$\qquad$&$\;\quad${$M_{B}$}$\qquad$&$\quad${$\Delta M_{\rm host}$}$\qquad$&$\quad${$\sigma_{\rm coh}$}$\qquad$ \\
\hline
&  &             &                                      SNLS\,+\,SDSS-II & &     \\
\hline
1. $R_{\rm h}=ct$  & $0.119\pm0.007$ &       $2.600\pm0.084$         &       $-18.932\pm0.021$       &       $-0.052\pm0.026$        & 0.106 (fixed)    \\

2. $w$CDM  & $0.121\pm0.008$ &       $2.631\pm0.087$         &       $-19.020\pm0.037$       &       $-0.051\pm0.028$        & 0.106 (fixed)          \\

3. $\Lambda$CDM & $0.121\pm0.008$ &       $2.631\pm0.086$         &       $-19.026\pm0.035$       &       $-0.051\pm0.028$     & 0.106 (fixed)         \\

4. $\Lambda$CDM ($k=0$) & $0.121\pm0.008$ &       $2.630\pm0.085$         &       $-19.040\pm0.030$       &       $-0.051\pm0.028$    & 0.106 (fixed)  \\

\hline
&  &             &                                    SNLS  & &    \\
\hline

5. $R_{\rm h}=ct$  & $0.108\pm0.014$ &       $2.290\pm0.153$         &       $-18.889\pm0.016$       &       $-0.043\pm0.023$          &$0.069\pm0.018$    \\

6. $w$CDM  & $0.114\pm0.014$ & $2.351\pm0.162$ & $-19.055\pm0.079$ & $-0.043\pm0.025$ & $0.056\pm 0.022$ \\

7. $\Lambda$CDM  & $0.114\pm0.014$ &       $2.351\pm0.162$         &       $-19.061\pm0.078$       &       $-0.043\pm0.025$       &$0.056\pm0.022$         \\

8. $\Lambda$CDM ($k=0$) & $0.113\pm0.014$ & $2.367\pm0.156$ & $-19.022\pm0.039$ & $-0.042\pm0.024$ &  $0.056\pm0.022$ \\
\hline
\end{tabular}
}
\caption{\label{tab:1}Optimized parameters for different cosmological models.}
\end{table*}

\begin{table*}
\centering
{
\tiny
\begin{tabular}{lccccc}
Model\qquad&\quad{$\Omega_{\rm m}$}\qquad&\quad{$\Omega_{\rm de}$}\qquad&$w_{\rm de}$&\quad{$\chi^{2}_{\rm dof}$}\qquad&\quad{BIC}\qquad \\
\hline
&  &             &                    SNLS\,+\,SDSS-II & &  \\
\hline
1. $R_{\rm h}=ct$  &       $\cdots$                 &  $\cdots$   & $\cdots$ & 1.04$\;\,$(\rm 609 dof) &  $-514.57$ \\

2. $w$CDM  & $0.203^{+0.137}_{-0.196}$          &  $0.445^{+0.265}_{-0.205}$     & $-0.956^{+0.276}_{-0.384}$ &  1.01$\;\,$(\rm 606 dof)  &  $-511.43$  \\

3. $\Lambda$CDM & $0.250\pm0.152$          &  $0.480\pm0.202$     & $-1$ (fixed)& 1.01$\;\,$(\rm 607 dof)  &  $-517.86$  \\

4. $\Lambda$CDM ($k=0$) & $0.353\pm0.045$          &  $1.0-\Omega_{\rm m}$     & $-1$ (fixed) & 1.01$\;\,$(\rm 608 dof)  &  $-523.70$  \\

\hline
&  &             &                   SNLS  & &  \\
\hline

5. $R_{\rm h}=ct$  &        $\cdots$                 &  $\cdots$   & $\cdots$ & 0.94$\;\,$(\rm 234 dof) &  $-177.65$ \\

6. $w$CDM &   $0.368\pm 0.120$   &  $0.806\pm0.310$   &   $-0.912^{+0.302}_{-0.446}$   &   0.98$\;\,$(\rm 231 dof)  &  $-169.99$ \\

7. $\Lambda$CDM  &   $0.453\pm0.130$               &  $0.869\pm0.336$         & $-1$ (fixed) & 0.97$\;\,$(\rm 232 dof)  &  $-173.61$ \\

8. $\Lambda$CDM ($k=0$) & $0.360\pm0.051$ &  $1.0-\Omega_{\rm m}$ & $-1$ (fixed) & 0.97$\;\,$(\rm 233 dof) &  $-178.20$ \\
\hline
\end{tabular}
}
\caption{\label{tab:2}Optimized parameters for different cosmological models (cont.)}
\end{table*}

But this is problematic for several reasons. First, one might 
expect that if a model is correct, it should fit either the SNLS, with
$\approx 250$ events distributed in redshift $0< z
< 1$, or the combined JLA sample with three times as many SNe
spread over a similar redshift range, comparably well, at least
qualitatively. The quality of the fit improves as the sample size
increases, but one would not expect the model selection to change
from one sample to the other because, in both cases, the SNe are
distributed across the region (near $z\approx 0.6$) where the
transition from deceleration to acceleration is thought to have
occurred.

Perhaps an indication of why there may be differences between the
analysis of SNLS on its own, versus the merged SNLS and
SDSS-II sample, is provided by the tendency of binned
residuals in both Figs.~\ref{fig:1} and~\ref{fig:2} to be slightly lower
at $z\lesssim 0.35$ than those at  $z\gtrsim 0.35$.  The average difference
is about $0.04$ magnitudes for $w$CDM and about $0.08$
magnitudes for $R_{\rm h}=ct$. On the other hand, for the 4~hightest
redshift bins, we find an average residual magnitude of $0.139$
for $R_{\rm  h}=ct$ and a slightly worse $0.143$ for $w$CDM\null.
Since these residuals do not exhibit any monotonic trend, the implication 
seems to be that there exists a systematic offset across $z\sim 0.35$.
One possible origin for this behavior could be that the calibration of the
SNLS and SDSS-II sources is not completely self-consistent
after all, and since the SDSS-II events occurred at
$0.05<z<0.4$, while the SNLS events were recorded at $0.2<z<1$, we may
simply be seeing the impact of an unresolved measurement offset in
their magnitude.  Notice, for example, that the implied magnitude
offset is comparable to the measured sample-wide `intrinsic'
dispersion $\sigma_{\rm coh}=\allowbreak 0.106\pm0.006$ (see
above). The offset may be slightly smaller for $w$CDM 
due to the additional free parameter that allows greater flexibility 
in shaping the best-fit curve. In both cases, however, the key point 
is that the offset appears to be independent of redshift above and 
below the crossover at $z\approx 0.35$.

Some other studies, e.g., ref.~\cite{Shafer:2015}, have 
reached different conclusions from those presented here. As we have discussed
extensively in this paper, although the statistical analysis of
Type~Ia SNe may be improved with the merger of disparate subsamples,
each subsample comes with its
own set of systematic and intrinsic uncertainties. Ref.~\cite{Shafer:2015} 
carried out the model selection using both the Union2.1 and JLA samples, 
following the conventional approach of minimizing an overall~$\chi^2$, though
with the constraint that $\chi^2_{\rm dof}=1$ for each subsample.  One
should not be surprised, therefore, that the $\chi^2_{\rm dof}$ for
the whole compilation is also close to~$1$. A correct statistical
approach, however, would estimate the unknown $\sigma_{\rm int}$'s
simultaneously with the model-specific and nuisance parameters
\cite{Kim:2011,Wei:2015}. One should therefore use~MLE\null.

More importantly, Union2.1 contains over 17 subsamples.  The total 
number of `nuisance' parameters is therefore~20, since all of the 
$\sigma_{\rm int}$'s should be re-estimated for each model for
a truly unbiased test. Moreover, 
the expressions used by ref.~\cite{Shafer:2015} to compute the
BIC are incorrect because the $\sigma_{\rm int}$'s themselves
are not known a~priori.  So the information criteria must be
calculated in terms of the likelihood function \cite{Wei:2015},
not~$\chi^2$. Finally, since ref.~\cite{Shafer:2015} used several previously
  optimized parameter values, those results are not unbiased
like the outcomes shown in lines 1--3 of Tables 1 and~2.

Comparing our results using the SNLS on its own and the
merged SNLS and SDSS-II sample shows that the model selection
using these Type~Ia SNe is still somewhat ambiguous. Our analysis 
has shown that the cross-calibration in the JLA may be imperfect, with 
a residual offset between the two samples of $\sim 0.08$ mag, comparable 
to the sample-wide dispersion $\sigma_{\rm coh}\sim 0.106$ mag inferred in
ref.~\cite{Betoule:2014}. Such an offset tends to favour models with a larger number 
of free parameters, with a greater flexibility in adjusting the shape
of their luminosity distance to fit the data. This may explain why 
$R_{\rm h}=ct$ is favoured over $w$CDM, but not always over $\Lambda$CDM.

An important goal of future work with Type~Ia SNe should be 
to improve the consistent cross-calibration of independent datasets,
along the lines initiated by ref.~\cite{Betoule:2014}, though with
even greater precision. If $R_{\rm h}=ct$ 
were to eventually become the cosmology favoured by the Type~Ia SN
data, such an outcome would be relevant to the growing tension 
between the predictions of an inflationary cosmology and the 
\emph{Planck} measurements. The Universe did not require an early 
period of inflated expansion to avoid the horizon problem 
in~$R_{\rm h}=ct$ \cite{Melia:2013b}, so the $R_{\rm h}=ct$ 
explanation for the uniformity of the physical conditions across 
the Universe may be more realistic than the currently 
held belief of an inflated expansion. 

\acknowledgments
We thank the
National Basic Research Program (`973' Program) of China (grants
2014\allowbreak{CB845800} and 2013\allowbreak{CB834900}), the National
Natural Science Foundation of China (grant nos.\ 11322328 and
11373068), the One-Hundred-Talents Program, the Youth Innovation
Promotion Association (2011231), the Strategic Priority Research
Program `The Emergence of Cosmological Structures' (grant
no.\ XDB09000000) of the Chinese Academy of Sciences, and the 
Natural Science Foundation of Jiangsu Province (grant no.\ BK20161096).

\end{document}